\documentclass{pasj01}
\draft 
\Received{$\langle$reception date$\rangle$}
\Accepted{$\langle$acception date$\rangle$}
\Published{$\langle$publication date$\rangle$}
\begin{document}

\title{How far actually is the Galactic Center IRS 13E3  from Sagittarius A$^{\ast}$?}
\author{Masato Tsuboi$^{1, 2}$, Yoshimi Kitamura$^1$,  Takahiro Tsutsumi$^3$, Ryosuke Miyawaki$^4$, Makoto Miyoshi$^5$ and Atsushi Miyazaki$^6$ }%
\altaffiltext{1}{Institute of Space and Astronautical Science, Japan Aerospace Exploration Agency,\\
3-1-1 Yoshinodai, Chuo-ku, Sagamihara, Kanagawa 252-5210, Japan }
\email{tsuboi@vsop.isas.jaxa.jp}
\altaffiltext{2}{Department of Astronomy, The University of Tokyo, Bunkyo, Tokyo 113-0033, Japan}
\altaffiltext{3}{National Radio Astronomy Observatory,  Socorro, NM 87801-0387, USA}
\altaffiltext{4}{College of Arts and Sciences, J.F. Oberlin University, Machida, Tokyo 194-0294, Japan}
\altaffiltext{5}{National Astronomical Observatory of Japan, Mitaka, Tokyo 181-8588, Japan}
\altaffiltext{6}{Japan Space Forum, Kanda-surugadai, Chiyoda-ku,Tokyo,101-0062, Japan}
\maketitle
\begin{abstract}
The Galactic Center IRS 13E cluster is a very intriguing IR object located at $\sim0.13$ pc from Sagittarius A$^\ast$ (Sgr A$^\ast$)  in projection distance. There are both arguments for and against the hypothesis that a dark mass like an intermediate mass black hole (IMBH) exists in the cluster. Recently we have detected the rotating ionized gas ring around IRS 13E3, which belongs to the cluster, in the H30$\alpha$ recombination line using ALMA. The enclosed mass is derived to be $M_{\mathrm{encl.}}\simeq2\times10^4$ $M_\odot$, which agrees with an IMBH and is barely less than the astrometric upper limit mass of the IMBH around  Sgr A$^\ast$. Because the limit mass depends on the true three-deminsional (3D) distance from  Sgr A$^\ast$, it is very important to determine it observationally. However,  the 3D distance is indefinite because it is hard to determine the line-of-sight (LOS) distance by usual methods. We would attempt to estimate the LOS distance by spectroscopic informations. The CH$_3$OH molecule is easily destroyed by cosmic ray around Sgr A$^{\ast}$. However, we detected a highly excited CH$_3$OH emission line in the ionized gas stream associated with  IRS 13E3. This indicates that IRS 13E3 is located at $r\gtrsim 0.4$ pc from  Sgr A$^{\ast}$.
\end{abstract}

\section{Introduction}
The Galactic Center IRS 13E cluster, which was thought to contain several WR and O stars in the early observations (e.g. \cite{Genzel1996}) is a very intriguing IR object located at $\sim0.13$ pc from Sagittarius A$^\ast$ (Sgr A$^\ast$)  in projection distance. The common direction and similar amplitude of the proper motions of the member stars suggest that  they are physically bound (e.g. \cite{Maillard} ) although the strong tidal force of Sgr A$^{\ast}$ should disrupt the cluster easily (e.g. \cite{Gerhard}). 
One of the possible speculations  is that a dark mass like an intermediate mass black hole (IMBH), $M\sim10^{4-5}$ $M_\odot$, in the cluster would prevent the cluster disruption (e.g. \cite{Maillard}). However,  the upper limit mass of the IMBH around Sgr A$^{\ast}$ by long-term VLBA astrometry (\cite{Reid}, \cite{Hansen}) is derived to be $M\lesssim3\times10^4$ $M_\odot$ if it is really located at the observed projection distance. Morever,  IR spectroscopic observations show that almost all the member stars in the cluster are not real stars but gas blobs. Therefore, such an IMBH is not always necessary (e.g.\cite{Schodel2009}, \cite{Genzel2010}).

However, the high concentration of the gas blobs itself is peculiar even in the Galactic center region.  We consider that the existence of the IMBH in the IRS 13E cluster is still an open question. We searched ionized gas with a very large velocity width and compactness in order to prove this hypothesis.
In our attempts, we had detected the ionized gas associated with IRS 13E3  in the H30$\alpha$ recombination line using ALMA (\cite{Tsuboi2017b} , \cite{Tsuboi2019}). The enclosed mass of IRS 13E3 is derived to be $M_{\mathrm{encl.}}\simeq2.4\times10^4 M_\odot$, which agrees with the mass being an IMBH \citep{Tsuboi2019}. The derived mass is barely less than the astrometric upper limit mass if the 3D distance of IRS 13E3 is nearly equal to the projection distance. 

The astrometric upper limit mass depends on the 3D distance from Sgr A$^{\ast}$. However,  the 3D distance is indefinite because the distance along the line-of-sight is hard to be derived by usual methods.  The 3D distance may be much larger than the projection distance.   We propose a new method to estimate the 3D distance from Sgr A$^{\ast}$.
Because the abundances of some molecules are considerably affected by the 3D distance, we would attempt to estimate the 3D distance by the spectroscopic informations of the molecules. First we searched the sign of the molecules in published data of the gas around the IRS 13E cluster.
Throughout this paper, we adopt 8.2 kpc as the distance to the Galactic center (e.g. \cite{Ghez}, \cite{Gillessen}, \cite{Schodel2009}, \cite{Boehle}): $1\arcsec$ corresponds to 0.04 pc or 8200 AU at the distance.

\begin{figure}
\begin{center}
\includegraphics[width=15cm, bb=0 0 1377.1 1207.2 ]{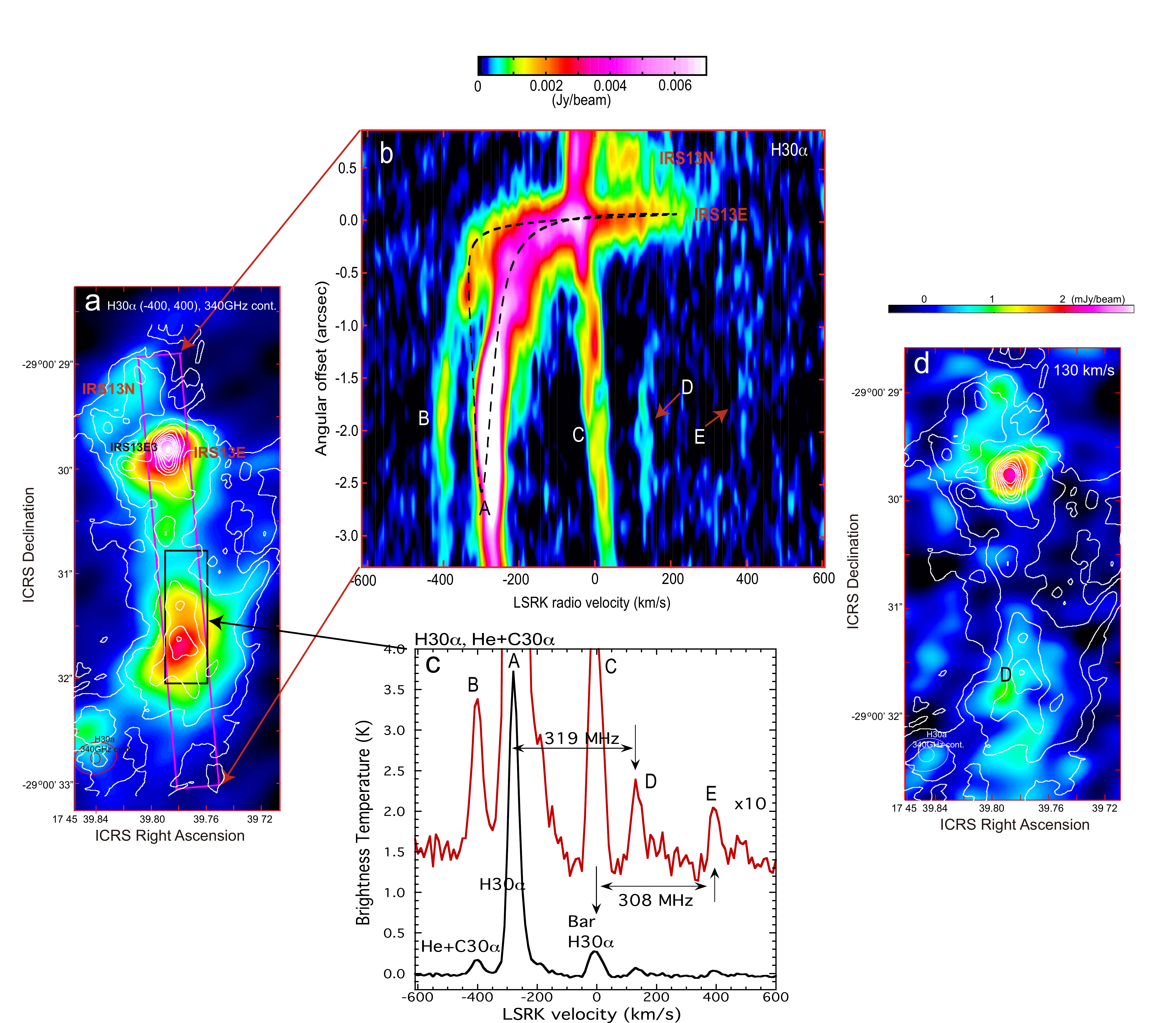}
\end{center}
\caption{{\bf a} Continuum map at 340 GHz (contours) and moment 0 map in the H30$\alpha$ recombination line (pseudocolor) of the IRS 13E cluster and the ionized gas stream depicted by a high-eccentricity Keplerian orbit (\cite{Tsuboi2017b}). The  integrated velocity range of the recombination line  is $V_\mathrm{LSR} = -400$ to $400$ km s$^{-1}$. The angular resolutions are $0\farcs107 \times 0\farcs101, PA=-78^\circ$ for the continuum map and $0\farcs41 \times 0\farcs30, PA=-77^\circ$  in FWHM  for the recombination line map, respectively. They are shown as the ovals at the lower left corner.
{\bf b} Position-velocity diagram in the H30$\alpha$ recombination line along a red rectangular shown in {\bf a}. The broken line shows the trajectory of the Keplerian orbit \citep{Tsuboi2017b}. 
{\bf c} Line profile in the H30$\alpha$ recombination line (black line) and ten-times magnification (red line). The sampling area is shown as a black rectangular shown in {\bf a}. 
{\bf d} Channel map in the H30$\alpha$ recombination line with $V_\mathrm{LSR} = 125$ to $135$ km s$^{-1}$ (pseudocolor). The contours show the continuum map at 340 GHz. }
\end{figure}

\section{Search for the molecular emission line in the gas flows around the IRS 13E cluster}
There are several published data with ALMA  of the gas associated with the IRS 13 cluster (e.g. \cite{Moser}, \cite{Tsuboi2018}).
First, we searched the sign of the easily found molecules like CO, CS, SiO and so on in the published data. However, we could not find the molecular gas (see also \cite{Tsuboi2017}) although some authors have reported the detections of the molecular lines in the region within the circum-nuclear molecular disk (CND)  (e.g. \cite{Moser},\cite{Yusef-Zadeh2017b}, \cite{Hsieh}).
The molecular gas would have a similar velocity feature in the PV diagram to that of the ionized gas if they are physically mixed in the same region.  If so, the use of the velocity feature of the ionized gas as a template should be useful to find the sign of the molecular emission from the IRS 13 cluster.

Figure 1b is the position-velocity (PV) diagram along a red rectangular shown in Figure 1a.  A feature labeled A is the ionized gas associated with IRS 13E3 in the H30$\alpha$ recombination line ($\nu_{\mathrm{rest}}= 231900.928$ MHz).  
While a feature labeled B is the counterpart in the He+C30$\alpha$ recombination line (He30$\alpha$; $\nu_{\mathrm{rest}}= 231995.428$ MHz, C30$\alpha$; $\nu_{\mathrm{rest}}= 232016.636$ MHz). In addition, a feature labeled C is the ionized gas belonging to the "Bar"  in the H30$\alpha$ recombination line, which is an extended part of the "Galactic center mini-spiral"(GCMS). Finally, we detected a similar velocity feature, a feature labeled D, in the PV diagram to that of the ionized gas associated with IRS 13E3  in the H30$\alpha$ recombination line. The line profiles of the area (a black rectangular shown in Figure 1a) are shown as Figure 1c. The line profiles of the features A, B, C and D are clearly identified as independent peaks.  The frequency difference between the Gaussian-fit central frequencies of the features A and D is about $\Delta \nu \sim 318.7$ MHz.  This corresponds to  the velocity difference of $ \Delta v\sim412.3$ km s$^{-1}$. The rest frequency of the feature D is expected to be $\nu_{\mathrm{rest}}= 231582.2$ MHz.  The candidate transitions corresponding to the rest frequency are summarized in Table 1.  The frequency difference and velocity difference from the feature A  of the feature D are also summarized in Table 1. 

\begin{table}
  \caption{Candidate transitions corresponding to the feature D.  }
  \label{tab:first}
 \begin{center}
    \begin{tabular}{cccccc}
 \hline   \hline
Name & transition &Rest frequency& Frequency  difference &Velocity difference&Remarks\\
& &$\nu_{\mathrm{rest}}$[MHz]&$\Delta \nu$ [MHz]&$\Delta V$[km s$^{-1}$]&\\
\hline
Hydrogen atom&H30$\alpha$&$231900.928$&-- &--&\\
\hline
t-CH$_3$CH$_2$OH&21(5,17)-21(4,18) &$231558.513^\ast$&$-342.415^\parallel$&443.0$^\parallel$& $T_{\mathrm{R}}=1.5$ K at Sgr B2 (N)$^\ddagger$ \\
t-CH$_3$CH$_2$OH&21(5,16)-20(4,17) &$231560.877^\ast$&$-340.051^\parallel$&439.9$^\parallel$&--\\
CH$_3$OH $v=1$&17(6,11)-17(7,10)&$231589.972^\natural$&$-310.956^\parallel$&402.3$^\parallel$& has not been observed\\
CH$_3^{13}$CH$_2$CN&26(3,24)-25(3,23)&$231590.269^\ast$&$-310.659^\parallel$&401.9$^\parallel$&$T_{\mathrm{R}}=0.6$ K at OriMC-1$^\dagger$ \\
CH$_3$CHO&12(3,10)-11(3,9)A$_{++}$&$231595.269^\ast$&$-305.659^\parallel$&395.4$^\parallel$&$T_{\mathrm{R}}=6.0$ K at Sgr B2 (N)$^\ddagger$ \\
CH$_3$CH$_2^{13}$CN&26(3,24)-25(3,23)&$231646.312^\ast$&$-254.616^\parallel$&329.4$^\parallel$&$T_{\mathrm{R}}=0.3$ K at OriMC-1$^\dagger$ \\
CH$_3$OCHO&27(3,25)-27(2,26)A&$231657.892^\ast$&$-243.036^\parallel$&314.4$^\parallel$&$T_{\mathrm{R}}=2.4$ K at Sgr B2 (N) $^\ddagger$ \\
$^{13}$CH$_3$CH$_2$CN&26(3,24)-25(3,23)&$231662.964^\ast$&$-237.964^\parallel$& 307.8$^\parallel$&$T_{\mathrm{R}}=0.4$ K at OriMC-1$^\dagger$ \\
CH$_3$CHO&12(-3,10)-11(-3,9)E&$231748.722^\ast$&$-152.206^\parallel$&196.9$^\parallel$&$T_{\mathrm{R}}=6.2$ K at Sgr B2 (N)$^\ddagger$ \\
\hline
Feature D&--&231582.2&$-318.7^\parallel$& $412.3^\parallel$ \\
Feature E&--&231592.5&$-308.9^\sharp$& $398.9^\sharp$ \\
\hline
    \end{tabular}
 \end{center}
$^\ast$ https://physics.nist.gov/cgi-bin/micro/table5/start.pl. $^\dagger$  \cite{Demyk}. $^\ddagger$ \cite{Nummelin}. $^\natural$\cite{Li-Hong}. $^\parallel$ from the feature A. $^\sharp$ from the feature C.
\clearpage
\end{table}

There are three transitions  of CH$_3$OH $v=1$, 17(6,11)-17(7,10), CH$_3^{13}$CH$_2$CN 26(3,24)-25(3,23), and 
CH$_3$CHO 12(3,10)-11(3,9)A$_{++}$ within $ \Delta V=\pm25$ km s$^{-1}$ from the velocity of the feature D, $V_{\mathrm{LSR}}\sim134$  km s$^{-1}$ (see Figure 1c). These are the candidate transitions of  the feature  D because features A and D have wide FWHM velocity widths of $\Delta V_{\mathrm{FWHM}}=50$ and $46$ km s$^{-1}$, respectively: the half width of the feature A is used as the tolerance mentioned above.
$^{13}$CH$_3$CH$_2$CN, CH$_3^{13}$CH$_2$CN, and CH$_3$CH$_2^{13}$CN molecules are isotope-substituted molecules of CH$_3$CH$_2$CN. These molecules are  in a family formed so that Carbon atoms, C, located at different positions are substituted by Carbon isotope atoms, $^{13}$C. The intensities of these emission lines are observed to be similar in warm and dense molecular clouds, for example, in Orion Molecular Cloud (e.g. \cite{Demyk}). Therefore, if the feature D is the CH$_3^{13}$CH$_2$CN 26(3,24)-25(3,23) emission line, the other emission lines should be also observed in some degree around $V_{\mathrm{LSR}}\sim30$  and $\sim50$ km s$^{-1}$. However, there is no sign in Figure 1c. The transition presumably is not  the feature D. 

Next we discuss the third candidate transition. CH$_3$CHO molecule has another transition, CH$_3$CHO 12(-3,10)-11(-3,9)E ($\nu_{\mathrm{rest}}=231748.722$ MHz) in the frequency range, which corresponds to the velocity difference of $\Delta V=196.9$ km s$^{-1}$ from the feature A. Because these transition parameters are similar, the intensities of these emission lines are observed to be similar in warm and dense molecular clouds, for example, in Sgr B2(N) (e.g. \cite{Nummelin}).  Therefore, if the feature D is the transition of the CH$_3$CHO molecule, the CH$_3$CHO 12(-3,10)-11(-3,9)E emission line should be also observed in some degree around $V_{\mathrm{LSR}}\sim-80$  km s$^{-1}$. However, there is no sign in Figure 1c. The transition presumably is not the feature D.

Finally, it is most probable that we detect the vibrationally excited CH$_3$OH emission line as the counterpart of the feature D. If this is the case, we have the first detection of the CH$_3$OH $v=1$, 17(6,11)-17(7,10) emission line ($\nu_{\mathrm{rest}}=231589.972$ MHz, \cite{Li-Hong}).
Figure 1d shows the feature D in the channel map of $V_{\mathrm{LSR}}=130$  km s$^{-1}$ referencing to  the H30$\alpha$ recombination line, which corresponds to $V_{\mathrm{LSR}}\sim-270$  km s$^{-1}$ referencing to  the CH$_3$OH $v=1$ emission line.  The velocity width is $\Delta V=10$  km s$^{-1}$. The feature can be identified up to  $\delta_{\rm ICRS}=-29^\circ00\arcmin31\arcsec$ along
the ionized gas stream depicted by a high-eccentricity Keplerian orbit (\cite{Tsuboi2017b}). 

The CH$_3$OH molecule is reported to be destroyed easily through the cosmic-ray photodissociation reaction; CH$_3$OH$+c.r.\to$H$_2$CO+H$_2$ (e.g. \cite{Harada}).   Because the cosmic ray intensity is thought to increase  significantly with approaching to Sgr A$^\ast$, CH$_3$OH molecule does not survive within some limit radius.
This is probably why the CH$_3$OH emission of the CND is very weak (see fig. 1e  in \cite{Tsuboi2018}). 
The limit radius is expected to be at least $r\sim 1$ pc from the projection distance of the innermost component of the CND in the CH$_3$OH emission line (\cite{Tsuboi2018}).
Therefore the detection of the CH$_3$OH emission line in the gas streamer suggests that the streamer is located at the 3D distance of $r\gtrsim1$ pc from Sgr A$^{\ast}$. 
Because the gas streamer is associated physically with IRS 13E3, IRS 13E3 is also located at a similar 3D distance from Sgr A$^{\ast}$, which is farther than the projection distance, $r_{\mathrm{p}}\sim 0.13$ pc.
If so,  the astrometric upper limit mass becomes larger and increases the possibility that an IMBH exists in the IRS 13E cluster. 

In addition, there is also a faint feature labeled E at $V_{\mathrm{LSR}}=399$  km s$^{-1}$ in figure 1c.  Because the frequency difference between the Gaussian-fit central frequencies of the features C and E is about $\Delta \nu \sim 308.9$ MHz or the velocity difference is $ \Delta V\sim398.9$ km s$^{-1}$, this is presumably the CH$_3$OH $v=1$, 17(6,11)-17(7,10) emission line corresponding to the feature C at $V_{\mathrm{LSR}}\sim -5$  km s$^{-1}$.  As mentioned above, the feature C corresponds to  the "Bar" of the GCMS.  The "Bar" is thought to obey a nearly pole-on high-eccentricity Keplerian orbit around Sgr A$^{\ast}$ (\cite{Tsuboi2017}),  and the part with $V_{\mathrm{LSR}}\sim 0$  km s$^{-1}$ corresponds to the vicinity of the apoastron of the orbit. The 3D distance from Sgr A$^{\ast}$ at the apoastron was estimated to be $a(1+e)\sim0.9$ pc. The 3D distance is consistent with the detection of the CH$_3$OH $v=1$, 17(6,11)-17(7,10) emission line there.

\begin{figure}
\begin{center}
\includegraphics[width=16cm, bb=0 0 1082.33 813.32 ]{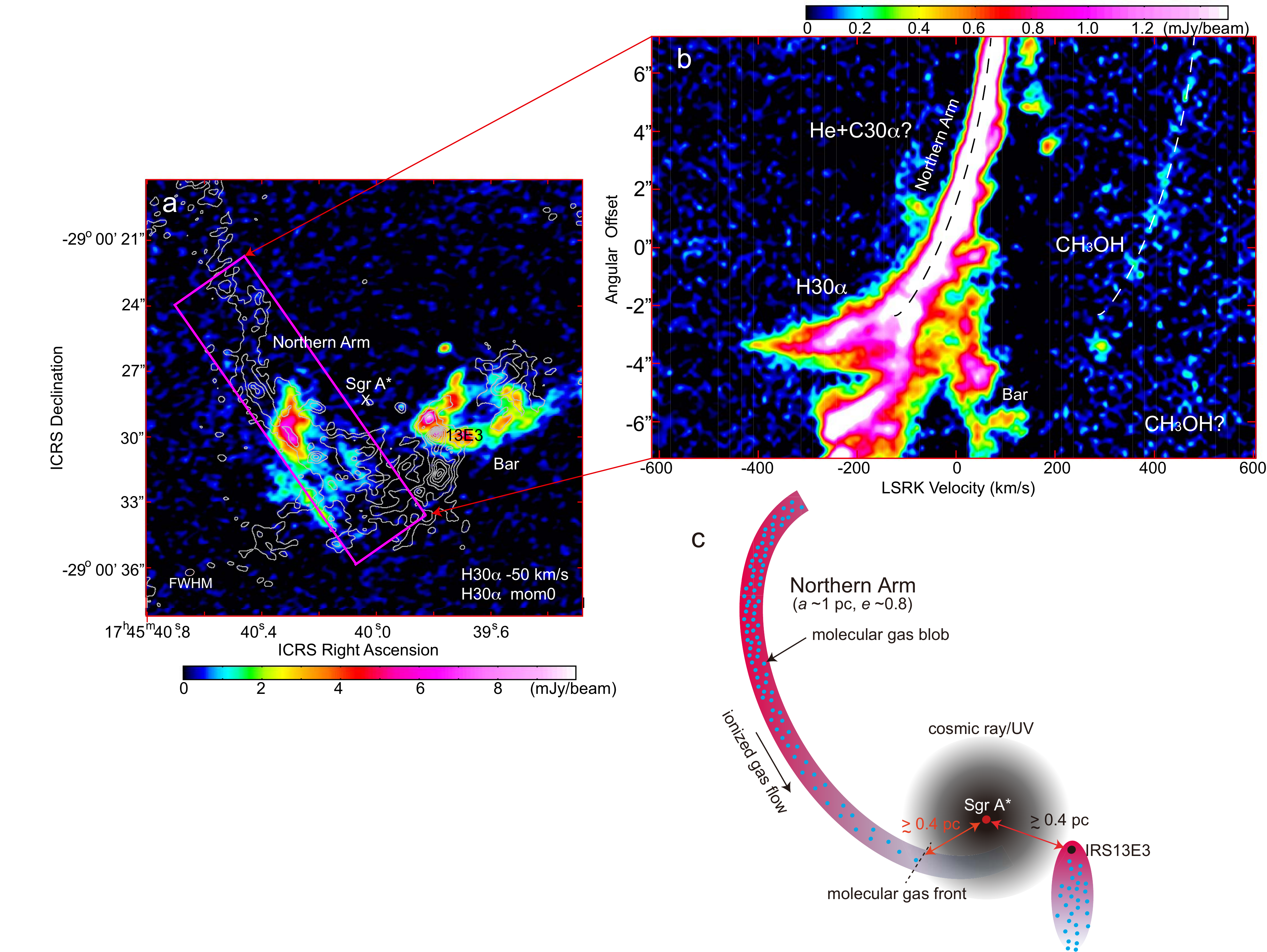}
\end{center}
\caption{{\bf a} Moment 0 map in the H30$\alpha$ recombination line with the integrated velocity range of $V_\mathrm{LSR} = -400$ to $400$ km s$^{-1}$ (contours) and channel map  of $V_\mathrm{LSR} = -55$ to $-45$ km s$^{-1}$ (pseudocolor).  The angular resolution is $0\farcs41 \times 0\farcs30, PA=-77^\circ$, which is shown as the oval at the lower left corner.  The white cross shows the position of Sgr A$^\ast$. {\bf b} Position-velocity diagram in the H30$\alpha$ recombination line along the Northern arm (a red rectangular shown in {\bf a}).  {\bf c} {\bf Schematic display of the positional relation among Sgr A$^\ast$, IRS 13E3, and the Northern Arm.}}
\end{figure}

\section{Vibrationally excited CH$_3$OH emission in the Northern arm}\label{sec:ODR}
We searched another example of the appearance and disappearance of the CH$_3$OH $v=1$  emission line in the GCMS.
Figure 2a shows the moment 0 map in the H30$\alpha$ recombination line with the integrated velocity range  of $V_\mathrm{LSR} = -400$ to $400$ km s$^{-1}$ (contours) and the channel map of $V_\mathrm{LSR} = -55$ to $-45$ km s$^{-1}$ (pseudocolor). 
Figure 2b is the PV diagram in the H30$\alpha$ recombination line along the Northern arm (NA) of the GCMS (a red rectangular shown in Figure 2a). The Northern arm is appeared as a long curved ridge in the diagram (a black broken line in Figure 2b). 
There may be a faint feature corresponding to He+C30$\alpha$ recombination line of the NA. 
As mentioned previously, the molecular gas would have a similar velocity feature in the PV diagram to that of the ionized gas if they are physically related to each other in the same region. Note that a long but faint curved ridge is identified  in the diagram  (a white broken line in Figure 2b).
 Because the velocity feature of the ridge is similar to the feature in the H30$\alpha$ recombination line and the velocity difference of $ \Delta V\sim400$ km s$^{-1}$ between the two ridges is the same as that discussed in the previous section, this probably corresponds to the  CH$_3$OH $v=1$  emission line. 

The ridge in the CH$_3$OH $v=1$  emission line is disappeared around the angular offset of $\Delta \theta \sim -1\arcsec$.  The apparent velocity at the disappearance point is $V_\mathrm{LSR} \sim 350$  km s$^{-1}$.   This corresponds to the ionized gas component with $V_\mathrm{LSR} \sim -50$  km s$^{-1}$ (see Figure 2a). Although the projection distance  from  Sgr A$^\ast$ of the disappearance point is $\theta_{\mathrm{p}} \sim -3\arcsec$ or $r_{\mathrm{p}} \sim 0.1$ pc, the 3D distance may be larger than the projection distance because this component is located at just before the intense acceleration from $V_\mathrm{LSR} \sim -50$   to $ \lesssim -200$ km s$^{-1}$.  Assuming the trajectory of the NA is described by a Keplerian orbit with the semi-major axis of $a\sim1$ pc, the eccentricity of $e\sim0.8$, and the inclination angle of $i\sim 140^\circ$ (\cite{Zhao2009}), the 3D distance of the disappearance point (molecular gas front) is estimated to be $r \sim 0.4$ pc (please see figure 2c). The NA is rapidly approaching to Sgr A$^\ast$ although the CND is rotating in a circular orbit around it. The elapse time moving from $r \sim 1$ pc to $r \sim 0.4$ pc is $\sim(1-2)\times10^3$ yr. The gradual decay would be why CH$_3$OH molecule is survived up to $r \sim 0.4$ pc.
It is assumed here that IRS 13E3 is located at $r \gtrsim 0.4$ pc. Nevertheless the astrometric upper limit mass increases up to $\sim 4\times10^4  M_\odot$ (\cite{Reid}, \cite{Hansen}), which is fairly larger than the estimated mass of IRS 13E3 \citep{Tsuboi2019}.

\begin{ack} 
This work is supported in part by the Grant-in-Aid from the Ministry of Eduction, Sports, Science and Technology (MEXT) of Japan, No.19K03939. The National Radio Astronomy Observatory (NRAO) is a facility of the National Science Foundation operated under cooperative agreement by Associated Universities, Inc. USA.  ALMA is a partnership of ESO (representing its member states), NSF (USA) and NINS (Japan), together with NRC (Canada), NSC and ASIAA (Taiwan), and KASI (Republic of Korea), in cooperation with the Republic of Chile. The Joint ALMA Observatory (JAO) is operated by ESO, AUI/NRAO and NAOJ. This paper makes use of the following ALMA data: ADS/JAO.ALMA\#2015.1. 01080.S and ALMA\#2017.1.00503.S.  

\end{ack}

\clearpage

\end{document}